\documentclass[12pt]{iopart}
\begin{document}

\title[Fluctuation-dissipation
for the Ising-Glauber model]{Fluctuation-dissipation relation
for the Ising-Glauber model with arbitrary exchange couplings}

\author{Christophe Chatelain\dag}
\address{\dag\ Laboratoire de Physique des Mat\'eriaux,
Universit\'e Henri Poincar\'e Nancy I,
BP~239, Boulevard des aiguillettes,
F-54506 Vand{\oe}uvre l\`es Nancy Cedex,
France}
\ead{chatelai@lpm.u-nancy.fr}

\begin{abstract}
We derive an exact expression of the response function to an infinitesimal
magnetic field for the Ising-Glauber model with arbitrary exchange couplings.
The result is expressed in terms of thermodynamic averages and does not
depend on initial conditions or dimension of space. The comparison with
the equilibrium case gives some understanding on the way the
fluctuation-dissipation theorem is violated out-of-equilibrium.
\end{abstract}

\submitto{\JPA}
\maketitle

\section{Introduction}
The knowledge about out-of-equilibrium processes is far from being as advanced
as for systems at equilibrium. In particular, the fluctuation-dissipation
theorem (FDT) which holds at equilibrium is known to be violated
out-of-equilibrium. This theorem states that at equilibrium the response
function $R_{\rm eq}(t,s)$ at time $t$ to an infinitesimal field branched to
the system at time $s$ is related to the time-derivative of the two-time
autocorrelation function $C_{\rm eq}(t,s)$:
\begin{equation}
  R_{\rm eq}(t,s)=\beta{\partial C_{\rm eq}(t,s)\over \partial s}.
  \label{eq0}
\end{equation}
In the Ising case, the response function reads $R(t,s)
={\delta\langle\sigma_i(t)\rangle\over\delta h_i(s)}$ and the
correlation function $C(t,s)=\langle\sigma_i(t)\sigma_i(s)\rangle$.
Based on a mean-field study of spin-glasses,
Cugliandolo \etal~\cite{Cugliandolo93, Cugliandolo94} have conjectured
that for asymptotically large times the FDT has to be corrected by a
multiplicative factor $X(t,s)$ which moreover depends on time only through
the correlation function:
\begin{equation}
  R(t,s)=\beta X(C(t,s)){\partial C(t,s)\over \partial s}.
  \label{eq-1}
\end{equation}
Exact results have been obtained for the
ferromagnetic Ising chain~\cite{Godreche00, Lippiello00} that confirm this
conjecture. The violation ratio $X(t,s)$ has been also numerically
computed for many systems: 2d and 3d-Ising ferromagnets~\cite{Barrat98},
3d Edwards-Anderson model~\cite{Franz95}, 3d and 4d Gaussian Ising spin
glasses~\cite{Marinari98}, 2d Ising ferromagnet with dipolar
interactions~\cite{Stariolo99},~$\ldots$
The response function has been exactly calculated in the case of
the Ising chain~\cite{Glauber63} but the result is not written in terms
of thermodynamic averages making difficult any conjecture on the way the
FDT is violated.

  We present an analytic study of the Glauber dynamics of the Ising model
with arbitrary exchange couplings. The response function to an infinitesimal
magnetic field is first calculated at equilibrium and the compatibility with
the fluctuation-dissipation theorem is verified. In the last section, the
response function is derived out-of-equilibrium and the expression is
compared to the equilibrium equivalent.

\section{Model, dynamics and general response function}
\subsection{The model and its dynamics}
We consider a classical Ising model whose degrees of freedom are scalar
variables $\sigma_i=\pm 1$ located on the nodes of a $d$-dimensional lattice.
Let us denote $\wp(\{\sigma\},t)$
the probability to observe the system in the state $\{\sigma\}$ at time $t$.
The dynamics of the system is supposed to be governed by the
master equation:
\begin{equation}
  \fl{\partial \wp(\{\sigma\},t)\over \partial t}
  =\sum_{\{\sigma'\}} \Big[
    \wp(\{\sigma'\},t)W(\{\sigma'\}\rightarrow \{\sigma\})
    -\wp(\{\sigma\},t)W(\{\sigma\}\rightarrow \{\sigma'\})\Big]
  \label{eq1}
\end{equation}
where $W(\{\sigma\}\rightarrow \{\sigma'\})$ is the transition rate from the
state $\{\sigma\}$ to the state $\{\sigma'\}$ per unit of time.
The normation is written
$\sum_{\{\sigma'\}}W(\{\sigma\}\rightarrow \{\sigma'\})=1$.
The transition rates are defined by the stationarity condition
\begin{equation}
  \sum_{\{\sigma'\}}\Big[
    \wp_{\rm eq}(\{\sigma'\})W(\{\sigma'\}\rightarrow \{\sigma\})
    -\wp_{\rm eq}(\{\sigma\})W(\{\sigma\}\rightarrow \{\sigma'\})
    \Big]=0
  \label{eq4}
\end{equation}
where $\wp_{\rm eq}(\{\sigma\})$ is the equilibrium probability distribution
which reads for the Ising model with general exchange-couplings:
\begin{equation}
  \wp_{\rm eq}(\{\sigma\})
  ={1\over {\cal Z}}e^{-\beta{\cal H}(\{\sigma\})}
  ={1\over {\cal Z}}e^{-\beta\sum_{k,l<k} J_{kl}\sigma_k\sigma_l}.
  \label{eq13}
\end{equation}
Equation (\ref{eq4}) is satisfied when the detailed balance holds:
\begin{equation}
  \wp_{\rm eq}(\{\sigma'\})W(\{\sigma'\}\rightarrow \{\sigma\})
  =\wp_{\rm eq}(\{\sigma\})W(\{\sigma\}\rightarrow \{\sigma'\}).
  \label{eq5}
\end{equation}
This last unnecessary but sufficient condition is fulfilled by
the heat-bath single-spin flip dynamics defined by the
following transition rates:
\begin{eqnarray}
  W(\{\sigma\}\rightarrow \{\sigma'\})
  &=\left[\prod_{l\ne k}\delta_{\sigma_l,\sigma'_l}\right]
  {\sum_\sigma \delta_{\sigma'_k,\sigma}
    \wp_{\rm eq}(\{\sigma_l\}_{l\ne k},\sigma_k=\sigma)
    \over\sum_\sigma \wp_{\rm eq}(\{\sigma_l\}_{l\ne k},\sigma_k=\sigma)}
  \nonumber\\
  &=\left[\prod_{l\ne k}\delta_{\sigma_l,\sigma'_l}\right]
  {e^{-\beta\sum_{l\ne k} J_{kl}\sigma'_l\sigma'_k}\over
  \sum_{\sigma=\pm 1} e^{-\beta\sum_{l\ne k} J_{kl}\sigma'_l\sigma}}.
  \label{eq6}
\end{eqnarray}
Only the single-spin flip $\sigma_k\rightarrow \sigma'_k$ is allowed.
The product of Kronecker deltas ensures that all other spins are not
modified during the transition. The spin $\sigma_k$ takes after the transition
the new value $\sigma'_k$ chosen
according to the equilibrium probability distribution
$\wp_{\rm eq}(\{\sigma\})$. All spins are successively updated during the
evolution of the system. These transition rates reduce to
Glauber's ones~\cite{Glauber63} in the case of the Ising chain.

\subsection{The response function}
A magnetic field $h_i$ is coupled to the spin $\sigma_i$ between the times
$s$ and $s+\Delta s$ where $\Delta s$ is supposed to be small. Between these
instants, the transition rates are changed to
$W_h(\{\sigma\}\rightarrow \{\sigma'\})$ satisfying equation
(\ref{eq6}) with an additional term $\beta h_i\sigma_i$ in the
Hamiltonian of the equilibrium probability distribution (\ref{eq13}):
\begin{equation}
  W_h(\{\sigma\}\rightarrow \{\sigma'\})
  =\left[\prod_{l\ne k}\delta_{\sigma_l,\sigma'_l}\right]
  {e^{-\beta\left[\sum_{l\ne k} J_{kl}\sigma'_l\sigma'_k
      -h_i\sigma'_k\delta_{k,i}\right]}
  \over\sum_{\sigma=\pm 1} e^{-\beta\left[
      \sum_{l\ne k} J_{kl}\sigma'_l\sigma-h_i\sigma\delta_{k,i}\right]}}.
\label{eq7}
\end{equation}
The transition rates are identical to the case $h_i=0$ apart from the
single-spin flip involving the spin $\sigma_i$.
Using the Bayes relation and the master equation (\ref{eq1}), the average
of the spin $\sigma_i$ at time $t$ can be expanded to lowest order
in $\Delta s$ under the following form:
\begin{eqnarray}
  \langle\sigma_i(t)\rangle
  &=\sum_{\{\sigma\}}\sigma_i\ \!\wp(\{\sigma\},t)
  \nonumber\\
  &=\sum_{\{\sigma\},\{\sigma'\}}\sigma_i\ \!
  \wp(\{\sigma\},t|\{\sigma'\},s+\Delta s)\wp(\{\sigma'\},s+\Delta s)
  \nonumber\\
  &=\sum_{\{\sigma\},\{\sigma'\}}\sigma_i\ \!
  \wp(\{\sigma\},t|\{\sigma'\},s+\Delta s)
  \Big[(1-\Delta s)\wp(\{\sigma'\},s)\Big.
  \nonumber\\
  &\hskip 100pt
  \Big.+\Delta s \sum_{\{\sigma''\}}\wp(\{\sigma''\},s)
  W_h(\{\sigma''\}\rightarrow \{\sigma'\})\Big].
  \label{eq2}
\end{eqnarray}
$W_h$ being the only quantity depending on the magnetic field in equation
(\ref{eq2}), the first term disappears after a derivative with respect
to the magnetic field. In the last term, the single-spin flip involving
the spin $\sigma_i$, i.e. the one coupled to the magnetic field, is the only
one that gives a non-vanishing contribution.
The integrated response function between the time $s$ and $s+\Delta s$ is
$\left[{\partial \langle\sigma_i(t)\rangle\over\partial h_i}
\right]_{h_i\rightarrow 0}$ and the limit $\Delta s\rightarrow 0$ leads to
the response function:
\begin{equation}
  \lim_{\Delta s\rightarrow 0}{1\over \Delta s}
  \left[{\partial \langle\sigma_i(t)\rangle\over\partial h_i}
  \right]_{h_i\rightarrow 0}
  =\lim_{\Delta s\rightarrow 0}
  {1\over \Delta s}\int^{s+\Delta s}_s R_{ii}(t,u)du
  =R_{ii}(t,s).
  \label{eq17}
\end{equation}
According to the definition of the response function (\ref{eq17}),
the calculation can be limited to the lowest order in $\Delta s$. As a
consequence, the conditional probability in equation (\ref{eq2})
can be replaced by $\wp(\{\sigma\},t|\{\sigma'\},s)$.

\section{Equilibrium fluctuation-dissipation relation}
\subsection{The response function}
If the system is in thermal equilibrium at time $s$, the probability
distribution $\wp(\{\sigma\},s)$ has to be replaced in equation (\ref{eq2})
by $\wp_{\rm eq}(\{\sigma\})$. Then the sum over ${\{\sigma''\}}$ can be
calculated:
\begin{eqnarray}
  \fl
  \sum_{\{\sigma''\}}&\wp_{\rm eq}(\{\sigma''\})
  W_h(\{\sigma''\}\rightarrow \{\sigma'\})
  \nonumber\\
  \fl
  &=\sum_{\{\sigma''\}}
  {1\over {\cal Z}}e^{-\beta\sum_{k,l<k} J_{kl}\sigma''_k\sigma''_l}\times
  \left[\prod_{j\ne i}\delta_{\sigma''_j,\sigma'_j}\right]
  {e^{-\beta\left[\sum_{j\ne i} J_{ij}\sigma'_i\sigma'_j-h_i\sigma'_i\right]}
    \over \sum_{\sigma^{(3)}_i=\pm 1}
    e^{-\beta\left[\sum_{j\ne i} J_{ij}\sigma^{(3)}_i\sigma'_j
        -h_i\sigma^{(3)}_i\right]}}
  \nonumber\\
  \fl
  &=\wp_{\rm eq}(\{\sigma'\})e^{\beta h_i\sigma'_i}
  {\cosh\left(\beta\sum_{j\ne i}J_{ij}\sigma'_j\right)
    \over\cosh\left(\beta\left[\sum_{j\ne i}J_{ij}\sigma'_j-h_i\right]\right)}.
  \label{eq18}
\end{eqnarray}
The derivative with respect to $h_i$ followed by the limit $h_i\rightarrow 0$
gives $\beta\wp_{\rm eq}(\{\sigma'\})\big[\sigma_i'
-\tanh\big(\beta\sum_{j\ne i}J_{ij}\sigma'_j\big)\big]$.
When replaced in equation (\ref{eq2}), one obtains
the following integrated response function
\begin{equation}
  \left[{\partial\langle\sigma_i(t)\rangle
      \over\partial h_i}\right]_{h_i\rightarrow 0}
  =\beta\Delta s\left[\langle\sigma_i(t)\sigma_i(s)\rangle
    -\langle\sigma_i(t)\tanh\Big(\beta\sum_{j\ne i}
    J_{ij}\sigma_j(s)\Big)\rangle\right]
  \label{eq19}
\end{equation}
and according to equation (\ref{eq17}) the response function reads
\begin{equation}
  R_{ii}(t,s)=\beta\left[\langle\sigma_i(t)\sigma_i(s)\rangle
    -\langle\sigma_i(t)\tanh\Big(\beta\sum_{j\ne i}
    J_{ij}\sigma_j(s)\Big)\rangle\right].
  \label{eq20}
\end{equation}

\subsection{Equivalence with the usual expression}
The equilibrium fluctuation-dissipation theorem states that the response
function is related to the time-derivative of the correlation function.
We will now show that the expression (\ref{eq20}) is compatible with this
statement.
At equilibrium, the correlation function is invariant under a
time-translation. As a consequence, one has the relation
\begin{equation}
  {\partial C_{ii}(t,s)\over\partial t}
  =-{\partial C_{ii}(t,s)\over\partial s}.
  \label{eq21}
\end{equation}
Moreover, it can be shown that the conditional probability
$\wp(\{\sigma\},t|\{\sigma'\},s)$ satisfies the master equation (\ref{eq1})
for the time-variable $t$ and a similar equation for $s$:
\begin{equation}
  \left(1-{\partial\over\partial s}\right)\wp(\{\sigma\},t|\{\sigma'\},s)
  =\sum_{\{\sigma''\}} \wp(\{\sigma\},t|\{\sigma''\},s)
  W(\{\sigma'\}\rightarrow \{\sigma''\}).
  \label{eq10}
\end{equation}
The time-derivative of the correlation function is then
\begin{eqnarray}
  \label{eq22}
  \fl
  {\partial C_{ii}(t,s)\over\partial t}
  &=\sum_{\{\sigma\},\{\sigma'\}} \sigma_i\sigma'_i
  {\partial \wp(\{\sigma\},t|\{\sigma'\},s)\over\partial t}
  \wp_{\rm eq}(\{\sigma'\})
  \\
  \fl
  &=-\sum_{\{\sigma\},\{\sigma'\}} \sigma_i\sigma'_i
  {\partial \wp(\{\sigma\},t|\{\sigma'\},s)\over\partial s}
  \wp_{\rm eq}(\{\sigma'\})
  \nonumber\\
  \fl
  &=-C_{ii}(t,s)+\sum_{\{\sigma\},\{\sigma'\},\{\sigma''\}}
  \sigma_i\sigma'_i\wp(\{\sigma\},t|\{\sigma''\},s)
  W(\{\sigma'\}\rightarrow \{\sigma''\})\wp_{\rm eq}(\{\sigma'\}).
  \nonumber
\end{eqnarray}
The sum over $\{\sigma''\}$ can be performed on the lines of equation
(\ref{eq18}) for the Ising-Glauber model. One obtains
\begin{eqnarray}
  \fl
  \sum_{\{\sigma'\}}\sigma_i'W(\{\sigma'\}\rightarrow \{\sigma''\})
  \wp_{\rm eq}(\{\sigma'\})
  &={1\over {\cal Z}}\sum_{\{\sigma'\}} \sigma_i'
  {\left[\prod_{j\ne i}\delta_{\sigma'_j,\sigma''_j}\right]
    e^{-\beta\sum_{j\ne i} J_{ij}\sigma''_i\sigma''_j}
    \over \sum_{\sigma_i^{(3)}=\pm 1}
    e^{-\beta\sum_{j\ne i} J_{ij}\sigma_i^{(3)}\sigma''_j}}
  e^{-\beta\sum_{k,l<k} J_{kl}\sigma'_k\sigma'_l}
  \nonumber\\
  \fl
  &=\wp_{\rm eq}(\{\sigma''\})
  \tanh\left(\beta\sum_{j\ne i}J_{ij}\sigma''_j\right).
  \label{eq23}
\end{eqnarray}
Replacing in (\ref{eq22}) and identifying with (\ref{eq20}),
one gets the expression of the equilibrium fluctuation-dissipation theorem
for the Ising-Glauber model:
\begin{equation}
  \fl
  R_{ii}(t,s)=\beta{\partial C_{ii}(t,s)\over\partial s}
  =\beta\left[\langle\sigma_i(t)\sigma_i(s)\rangle
    -\langle\sigma_i(t)\tanh\Big(\beta\sum_{j\ne i}
    J_{ij}\sigma_j(s)\Big)\rangle\right].
   \label{eq24}
\end{equation} 

\section{General fluctuation-dissipation relation}
Calculations have been made up to now within the hypothesis that the system
is at equilibrium at time $s$. We will now derive a relation valid even if the
system is far from equilibrium. We start with the
expansion (\ref{eq2}) of the average $\langle\sigma_i(t)\rangle$.
$W_h$ being the only quantity depending on the magnetic field, the
integrated response function between the time $s$ and $s+\Delta s$ is
to lowest order in $\Delta s$ 
\begin{equation}
  \fl\left[{\partial \langle\sigma_i(t)\rangle\over\partial h_i}
  \right]_{h_i\rightarrow 0}\!\!\!\!=\Delta s\!\!\!\!
  \sum_{\{\sigma\},\{\sigma'\},\atop\{\sigma''\}}\!\!
  \sigma_i\ \!\wp(\{\sigma\},t|\{\sigma'\},s)
  \left[{\partial W_h\over\partial h_i}(\{\sigma''\}\rightarrow \{\sigma'\})
    \right]_{h_i\rightarrow 0}\!\!\!\!\wp(\{\sigma''\},s).
  \label{eq3}
\end{equation}
The derivative of the transition rate $W_h$ defined by equation (\ref{eq7})
is easily taken and reads
\begin{equation}
  \fl
  \left[{\partial W_h\over\partial h_i}(\{\sigma''\}\rightarrow \{\sigma'\})
  \right]_{h_i\rightarrow 0}\!\!\!\!
  =\beta W(\{\sigma''\}\rightarrow \{\sigma'\})\left[
    \sigma'_i-\tanh\left(\beta\sum_{j\ne i} J_{ij}\sigma_j'\right)\right].
  \label{eq8}
\end{equation}
Combining (\ref{eq3}) and (\ref{eq8}), the integrated response function
is rewritten as
\begin{eqnarray}
  \fl
  \left[{\partial \langle\sigma_i(t)\rangle\over\partial h_i}
  \right]_{h_i\rightarrow 0}
  =&\Delta s\beta\sum_{\{\sigma\},\{\sigma'\},\atop\{\sigma''\}}
  \sigma_i\sigma'_i\ \!\wp(\{\sigma\},t|\{\sigma'\},s)
  W(\{\sigma''\}\rightarrow \{\sigma'\})\wp(\{\sigma''\},s)
  \nonumber\\
  \fl
  &-\Delta s\beta\sum_{\{\sigma\},\{\sigma'\},\atop
    \{\sigma''\}} \sigma_i\ \!\wp(\{\sigma\},t|\{\sigma'\},s)
  W(\{\sigma''\}\rightarrow \{\sigma'\})
  \nonumber\\
  \fl
  &\hskip 100pt\times\tanh\left(\beta\sum_{j\ne i} J_{ij}\sigma_j'\right)
  \wp(\{\sigma''\},s).
  \label{eq9}
\end{eqnarray}
In the first term of (\ref{eq9}), the sum over $\{\sigma''\}$ can be performed
using the master equation (\ref{eq1}) and the expression reorganised using
the fact that after the limit $h_i\rightarrow 0$, the conditional probability
becomes invariant under a time-translation and depends only on the
difference $t-s$. One obtains
\begin{eqnarray}
  \sum_{\{\sigma\},\{\sigma'\},\atop\{\sigma''\}}
  &\sigma_i\sigma'_i\ \!\wp(\{\sigma\},t|\{\sigma'\},s)
  W(\{\sigma''\}\rightarrow \{\sigma'\})\wp(\{\sigma''\},s)
  \nonumber\\
  &=\sum_{\{\sigma\},\{\sigma'\}}
  \sigma_i\sigma'_i\ \!\wp(\{\sigma\},t|\{\sigma'\},s)
  \left(1+{\partial\over\partial s}\right)\wp(\{\sigma'\},s)
  \nonumber\\
  &=C_{ii}(t,s)+{\partial C_{ii}(t,s)\over\partial s}
  -\sum_{\{\sigma\},\{\sigma'\}}
  \sigma_i\sigma'_i\ \!{\partial\wp\over\partial s}(\{\sigma\},t|\{\sigma'\},s)
  \wp(\{\sigma'\},s)
  \nonumber\\
  &=C_{ii}(t,s)+{\partial C_{ii}(t,s)\over\partial s}
  +{\partial C_{ii}(t,s)\over\partial t}.
   \label{eq25}
\end{eqnarray}
Similarly, in the second term of equation (\ref{eq9}) the sum over
$\{\sigma''\}$ can be performed using the master equation (\ref{eq1}) and
leads to
\begin{eqnarray}
  \fl
  \sum_{\{\sigma\},\{\sigma'\},\atop
    \{\sigma''\}} &\sigma_i\ \!\wp(\{\sigma\},t|\{\sigma'\},s)
  \tanh\left(\beta\sum_{j\ne i} J_{ij}\sigma_j'\right)
  W(\{\sigma''\}\rightarrow \{\sigma'\})\wp(\{\sigma''\},s)  
  \nonumber\\
  \fl
  &=\sum_{\{\sigma\},\{\sigma'\}} \sigma_i
  \ \!\wp(\{\sigma\},t|\{\sigma'\},s)
  \tanh\left(\beta\sum_{j\ne i} J_{ij}\sigma_j'\right)
  \left(1+{\partial\over\partial s}\right)\wp(\{\sigma'\},s)
  \nonumber\\
  \fl
  &=\left(1+{\partial\over\partial s}+{\partial\over\partial t}\right)
  \sum_{\{\sigma\},\{\sigma'\}} \sigma_i
  \ \!\wp(\{\sigma\},t|\{\sigma'\},s)
  \tanh\left(\beta\sum_{j\ne i} J_{ij}\sigma_j'\right)\wp(\{\sigma'\},s).
  \label{eq26}
\end{eqnarray}
Putting together equations (\ref{eq25}) and (\ref{eq26}) into (\ref{eq9}),
one can then evaluate the response function (\ref{eq17}):
\begin{equation}
  \fl
  R_{ii}(t,s)
  =\beta\left(1+{\partial\over\partial s}
    +{\partial\over\partial t}\right)\left[
    C_{ii}(t,s)-\langle\sigma_i(t)\tanh\left(\beta\sum_{j\ne i}
      J_{ij}\sigma_j(s)\right)\rangle\right].
  \label{eq29}
\end{equation}
The hyperbolic tangent is the equilibrium value of a single spin
$\sigma_i$ in the Weiss field created by all other spins at time $s$.
Moreover, the hyperbolic tangent being an odd function of its argument and since
$\sigma_i^2=1$, one can rewrite (\ref{eq29}) in the following form:
\begin{equation}
  \fl
  R_{ii}(t,s)
  =\beta\left(1+{\partial\over\partial s}
    +{\partial\over\partial t}\right)\left[
    C_{ii}(t,s)-\langle\sigma_i(t)\sigma_i(s)
    \tanh\Big(\beta\epsilon_i(s)\Big)\rangle\right].
  \label{eq30}
\end{equation}
where $\epsilon_i(s)=\sum_{j\ne i}J_{ij}\sigma_i(s)\sigma_j(s)$ is the
energy at site $i$.

\section{Conclusion}
The expression (\ref{eq29}) (or (\ref{eq30})) of the response function makes
apparent the way the FDT is violated. At equilibrium, all averages are invariant
under a time-translation, i.e. depend only on $t-s$ so the derivatives over
$s$ and $t$ cancel and at equilibrium the average in equation (\ref{eq29})
is equal to the time-derivative of the correlation function as shown 
in section 3.2.
The expression (\ref{eq29}) is very general. Since the response function
is expressed as thermodynamic averages, it does not depend on the initial
conditions. Moreover, it does not depend on the dimension of space or whether
or not, the lattice is regular. It applies to any configuration of exchange
couplings so for a disordered system, the average response function is just
the average of expression (\ref{eq29}) over all couplings configurations.
Equations (\ref{eq25}) and (\ref{eq26}) suggest
that the presence of the term $1+{\partial\over\partial s}+
{\partial\over\partial t}$ in the expression of the response function may be
a general feature of the response function out-of-equilibrium.
Exact results for other models are highly desirable.
The last term of (\ref{eq29}) whose equilibrium limit is the time-derivative
of the correlation function may also be evaluated perturbatively
for certain models. 

This work may also help to improve numerical simulations.
Up to now, the FDT has been numerically checked by calculating the
time-integrated response function. One can note that equation~(\ref{eq3})
can be used to calculate the response function directly during a Monte Carlo
simulation.

\ack
The laboratoire de Physique des Mat\'eriaux is Unit\'e Mixte de Recherche
CNRS number 7556. The author gratefully acknowledges Dragi Karevski,
Lo{\"\i}c Turban, Bertrand Berche and Alan Picone for critical reading of the
manuscript.


\section*{References}
\begin{thebibliography}{16}

\bibitem{Cugliandolo93} Cugliandolo L F and Kurchan J 1993
  {\it Phys. Rev. Lett} {\bf 71} 173
\bibitem{Cugliandolo94} Cugliandolo L F and Kurchan J 1994
  {\it J. Phys. A} {\bf 27} 5749
\bibitem{Godreche00} Godreche C and Luck J M 2000a
  {\it J.Phys. A} {\bf 33} 1151
\bibitem{Lippiello00} Lippiello E and Zannetti M 2000
  {\it Phys. Rev. E} {\bf 61} 3369
\bibitem{Barrat98} Barrat A 1998 {\it Phys. Rev. E} {\bf 57} 3629
\bibitem{Franz95} Franz S and Rieger R 1995
  {\it J. Stat. Phys.} {\bf 79} 749
\bibitem{Marinari98} Marinari E, Parisi G, Ricci-Tersenghi F, Ruiz-Lorenzo J
  1998 {\it J. Phys. A} {\bf 31} 2611
\bibitem{Stariolo99} Stariolo D A and Cannas S A 1999
  {\it Phys. Rev. B} {\bf 60} 3013
\bibitem{Glauber63} Glauber R J 1963 {\it J. Math. Phys} {\bf 4} 294 
\end{thebibliography}
\end{document}